\documentclass[aps,prb,reprint,superscriptaddress,amsmath,amssymb,longbibliography,floatfix]{revtex4-2}
\usepackage[T1]{fontenc}
\usepackage{lmodern}
\usepackage{microtype}
\usepackage{graphicx}
\usepackage{booktabs}
\usepackage{bm}
\usepackage{mathtools}
\usepackage{braket}
\usepackage{xcolor}
\usepackage{algorithm}
\usepackage{algpseudocode}
\usepackage{float}
\usepackage{subcaption}

\algrenewcommand{\algorithmicrequire}{\textbf{Input:}}
\algrenewcommand{\algorithmicensure}{\textbf{Output:}}

\usepackage[unicode,pdfencoding=auto]{hyperref}
\usepackage{orcidlink}
\hypersetup{
  colorlinks=true,
  linkcolor=blue!50!black,
  citecolor=blue!50!black,
  urlcolor=blue!50!black,
  pdftitle={Matrix-product-state GPLHR for non-Hermitian many-body eigenproblems},
  pdfauthor={Xingyao Wu; Yanbin Chen; Yu Wang; Zhangyu Yang}
}
\newcommand{\tc}{\mathrm{TC}}

\begin{document}
\floatstyle{ruled}
\newfloat{algorithm}{htbp}{loa}
\floatname{algorithm}{Algorithm}
\title{Matrix-product-state GPLHR for non-Hermitian many-body eigenproblems}
\input{authors}

\begin{abstract}
Computing multiple eigenstates of non-Hermitian many-body operators plays an important role in investigating excitation spectra, spectral properties, and relaxation dynamics in interacting and open quantum systems. However, this task is challenging because the eigenvectors are generally nonorthogonal and the Hilbert space dimension grows exponentially with system size. Here, we formulate the generalized preconditioned locally harmonic residual (GPLHR) method in a matrix product state (MPS) framework to efficiently compute multiple targeted right eigenstates whose eigenvalues lie near a specified shift in the complex plane. Successive randomized compression (SRC) is employed to accelerate operator applications. We test the method on transcorrelated and complex Hubbard chains, an interacting non-Hermitian Su–Schrieffer–Heeger (SSH) chain, and a Liouvillian model. The results demonstrate that GPLHR combined with MPS provides an efficient and accurate approach for computing multiple eigenstates of non-Hermitian many-body operators.
\end{abstract}
\maketitle
\raggedbottom
\section{Introduction}
\label{sec:introduction}

Computing several eigenstates of a non-Hermitian many-body operator is important for understanding excitation spectra,
nonreciprocal transport, and dissipative dynamics. Such problems arise in many non-Hermitian Hamiltonians and open
quantum systems. It is also attracting increasing interest in quantum chemistry and correlated lattice models, as the eigenstates may admit a more compact representation when part of the correlation is transferred from the wavefunction to the Hamiltonian, rendering the latter non-Hermitian~\cite{BoysHandy1969,CohenEtAl2019,BaiardiReiher2020,
BaiardiEtAl2022,LiaoEtAl2023}.

However, computing eigenstates of quantum many-body systems remains a central challenge in computational physics due to the so-called ``curse of dimensionality''. The exponential growth of the Hilbert-space dimension with system size limits exact methods, such as exact diagonalization (ED), to small system sizes~\cite{BaiEtAl2000,Saad2011}.
Non-Hermitian systems pose additional challenges, as their right eigenvectors can be strongly nonorthogonal and their eigenvalues can be highly sensitive to small perturbations~\cite{Trefethen1997}.
These properties make the accurate computation of multiple eigenstates more challenging. 

Fortunately, tensor network representations, e.g., matrix product states (MPS) and matrix product operators (MPO), offer compact representations of quantum states and operators~\cite{White1992,OstlundRommer1995,VerstraeteEtAl2008,
Schollwock2011}.
Recent developments in non-Hermitian DMRG further demonstrate the potential of tensor-network methods in obtaining the ground state of non-Hermitian quantum system~\cite{ZhongEtAl2025,TangEtAl2025}. However, even with MPS, accurately computing excited states remains challenging. Conventional excited-state DMRG often optimizes states sequentially, which becomes costly
when many states are required~\cite{BaiardiEtAl2022}.  Existing MPS eigensolvers use different strategies to target excited states. The shift-and-invert matrix product state (SIMPS) method targets an eigenstate near a chosen energy $\sigma$ by repeatedly applying an approximation to the shifted inverse $(H-\sigma I)^{-1}$, while DMRG[FEAST] constructs a spectral subspace from solutions of linear systems at several complex shifts~\cite{Yu2017,BaiardiFEAST2022}. Both approaches require repeated solutions of linear systems within the MPS representation, which can be computationally expensive. Although DMRG[FEAST] allows approximate solutions and improves the subspace through outer iterations, it still requires solutions at multiple shifts. MPS Lanczos methods~\cite{WangEtAl2025} avoid these shifted systems by building search spaces through Hamiltonian applications. However, this method focuses on low-lying states of Hermitian Hamiltonians, rather than states near a chosen point in a complex spectrum. In this work, we therefore propose an MPS formulation of the generalized preconditioned locally harmonic residual (GPLHR) method for non-Hermitian many-body eigenproblems. We demonstrate that this method can mitigate the aforementioned challenges, enabling the accurate and efficient computation of multiple eigenstates of non-Hermitian many-body operators within a chosen region of the complex spectrum. GPLHR is a block iterative eigensolver targeting multiple eigenstates near a specified shift in the complex plane~\cite{VecharynskiEtAl2016}. It constructs its search subspace through preconditioned residual expansions and harmonic projection, without requiring an exact shift-and-invert transformation. The iteration uses an orthonormal Schur basis to approximate the target invariant subspace, avoiding the numerical difficulties associated with strongly nonorthogonal eigenvectors. Combining GPLHR with MPS therefore offers a practical approach to computing multiple targeted eigenstates of large many-body systems with reduced storage requirements.

In this paper, we assess our method on non-Hermitian Hubbard chains, interacting  SSH chains, and a Liouvillian model. The benchmarks include a 128-site interacting SSH chain and the computation of twenty low-lying many-body states for a Hubbard chain, illustrating
both system-size scalability and accurate access to multiple eigenstates.  The rest of the paper is organized as follows.
Section~\ref{sec:mps_gplhr} introduces the GPLHR method, reviews the MPS representation, including the SRC method used for MPS compression, and presents the MPS-GPLHR method. Section~\ref{sec:models} introduces the models used in this paper and reports the benchmark results.

\section{Methods}
\label{sec:mps_gplhr}

We aim to compute the right eigenstates of a non-Hermitian Hamiltonian,
\begin{equation}
H\ket{\psi_i}=\lambda_i\ket{\psi_i},
\qquad H\neq H^\dagger,
\label{eq:right_eigenproblem}
\end{equation}
with eigenvalues near a prescribed shift $\sigma$. We first review the arithmetic GPLHR algorithm in its standard matrix-vector representation~\cite{VecharynskiEtAl2016}. We then briefly review the MPS formalism, and formulate the MPS-GPLHR method.

\subsection{GPLHR algorithm}
\label{sec:gplhr_algorithm}

\subsubsection{Schur residuals and search directions.}
\label{sec:gplhr_residual}
For non-Hermitian problems, individual right eigenvectors
can be strongly nonorthogonal to each other.
Collecting $b$ linearly independent target eigenvectors
into $X=[\psi_1,\ldots,\psi_b]$, the eigenvalue equations
can be written as
\begin{equation}
    HX=X\Lambda,
    \qquad
    \Lambda=\operatorname{diag}(\lambda_1,\ldots,\lambda_b).
\end{equation}
Thus, their span is invariant under the action of $H$.
GPLHR approximates this invariant subspace using an
orthonormal basis and subsequently recovers individual
eigenvectors as linear combinations of its columns.

Let $V=[v_1,\ldots,v_b]\in\mathbb C^{n\times b}$ be an
orthonormal basis spanning the exact target invariant
subspace, where $b$ is the number of desired eigenstates
and $n$ is the dimension of the Hilbert space under
consideration.
Following Ref.~\cite{VecharynskiEtAl2016}, we express $V$ through a partial generalized Schur decomposition of the matrix pair $(H,I)$, which takes the form
\begin{equation}
    HV=Q_{\mathrm S}R_A,
    \qquad
    V=Q_{\mathrm S}R_B,
\end{equation}
where $Q_{\mathrm S}\in\mathbb C^{n\times b}$ has
orthonormal columns and
$R_A,R_B\in\mathbb C^{b\times b}$ are upper triangular.
Here $Q_{\mathrm S}$ is an auxiliary Schur basis.
Eliminating it through the construction in
Ref.~\cite{VecharynskiEtAl2016} gives the $Q$-free relation
$HVM_B=VM_A$, with upper triangular factors $M_A,M_B$.

During the iteration, let $V=[v_1,\ldots,v_b]$ be
an orthonormal basis for the current approximation
to the target invariant subspace, with $V^\dagger V=I_b$.
The corresponding approximate relation is
\begin{equation}
    HVM_B\simeq VM_A,
    \label{eq:qfree_schur}
\end{equation}
where $M_A,M_B\in\mathbb C^{b\times b}$ are upper triangular.
For nonsingular $M_B$, this relation is equivalent to
$HV\simeq VS$, where $S=M_AM_B^{-1}$ is upper triangular.
Its diagonal entries are
$S_{jj}=(M_A)_{jj}/(M_B)_{jj}$ and provide approximations
to the target eigenvalues.
The symbol $\simeq$ means that the two sides are approximately equal, and their difference is measured by the Schur residual 
\begin{equation}
    R=HVM_B-VM_A.
    \label{eq:residual}
\end{equation}
A small residual indicates that the current basis and
triangular factors approximately satisfy the
invariant-subspace relation. Since the initial subspace generally provides
only an approximation to the target subspace, this residual
is nonzero. At each iteration, GPLHR evaluates the residual
and uses it to construct correction directions that expand
the trial subspace and improve the approximation.

The current basis $V$ spans only an approximate invariant subspace; we therefore seek a block correction $\delta V$
that moves the current subspace toward the desired invariant subspace.
Let $\delta V$ be the correction to the current basis $V$, and let
$\delta M_A$ and $\delta M_B$ be the corresponding corrections to the
triangular factors. We choose the correction to be orthogonal to the current subspace, $V^\dagger \delta V = 0$. Substituting these corrected quantities into Eq.~\eqref{eq:qfree_schur} gives
\begin{equation}
    H(V+\delta V)(M_B+\delta M_B)
    =(V+\delta V)(M_A+\delta M_A).
    \label{eq:corrected_schur}
\end{equation}
Expanding Eq.~\eqref{eq:corrected_schur}, neglecting
products of the correction terms, and using the previously
defined Schur residual $R$ gives
\begin{equation}
    H\delta V M_B-\delta V M_A
    +HV\delta M_B-V\delta M_A
    \simeq -R.
\end{equation}
Let $P_V = I - VV^\dagger$ be the orthogonal projector onto the orthogonal
complement of $V$. We then have
$P_V\delta V = \delta V$ and $P_VV = 0$.
Therefore, applying $P_V$ to the expanded equation gives
\begin{equation}
    P_VH\delta V M_B-\delta V M_A
    +P_VHV\delta M_B
    \simeq -P_VR.
    \label{eq:6}
\end{equation}
Also, for non-singular $M_B$, Eq.~\eqref{eq:residual} can be rewritten as
\begin{equation}HV=VM_AM_B^{-1}+RM_B^{-1}.
\end{equation}
Since $P_VV=0$, the $P_VHV\delta M_B$ term in Eq.~\eqref{eq:6}
becomes $P_VRM_B^{-1}\delta M_B$.
Near convergence, both the residual $R$ and the relative correction
$M_B^{-1}\delta M_B$ are small, so their product is of higher order.
Neglecting this term gives the equation
\begin{equation}
    (P_VHP_V)\delta V M_B-\delta V M_A
    =-\widetilde R,
\label{eq:block_correction}
\end{equation}
with $\widetilde R=P_VR$,
whose solution provides an approximation to the desired correction
$\delta V$ in Eq.~\eqref{eq:corrected_schur}. 

We write the left-hand side of the above equation as
a linear operator
\begin{equation}
    \mathcal L(D)=(P_VHP_V)DM_B-DM_A,
\end{equation}
subject to the constraint
\begin{equation}
    V^\dagger D=0.
\end{equation}
Then, Eq.~\eqref{eq:block_correction} becomes
\begin{equation}
    \mathcal L(\delta V)=-\widetilde R.
\end{equation}
If $\mathcal L$ is invertible on the
constrained space, the correction is
$\delta V=-\mathcal L^{-1}(\widetilde R)$.

However, solving this coupled matrix equation directly is generally expensive.
Therefore, we introduce a simple preconditioner $T_L$ to approximate
$\mathcal{L}^{-1}$ and use the projected residual to generate a correction subspace.
To construct the preconditioner, we follow the approach in Ref.~\cite{VecharynskiEtAl2016} and approximate $M_A$ and $M_B$ by $\sigma I$ and $I$, respectively, where $I$ is the $b\times b$ identity matrix. The inverse of the operator $\mathcal{L}(\cdot)$ is then approximated by the preconditioner $T_L = P_V T_\sigma P_V$, where $T_\sigma \approx (H-\sigma I)^{-1}$.

Applying $T_L$ to the projected residual and orthonormalizing the result gives the first correction block
\begin{equation}
    W=\operatorname{orth}\!\left(T_L\widetilde R\right).
\end{equation}
To enlarge the correction subspace, we generate additional blocks $S_1,\ldots,S_m$ through a preconditioned Arnoldi-like recurrence, where the input parameter $m$ determines the size of the correction subspace. Starting with $S_0=W$, for $\ell=1,\ldots,m$,
\begin{align}
    \widehat S_\ell
    &=T_L\!\left(HS_{\ell-1}M_B-S_{\ell-1}M_A\right),
    \label{eq:search_expansion}
    \\[4pt]
    S_\ell
    &=\operatorname{orth}\!\left[
    P_{S^\perp}^{(\ell-1)}
    (I-WW^\dagger)\widehat S_\ell
    \right].
    \label{eq:search_expansion_orth}
\end{align}
Here, $P_{S^\perp}^{(\ell-1)}$ is the orthogonal projector onto the orthogonal complement of the subspace spanned by $S_1,\ldots,S_{\ell-1}$:
\begin{equation}
    P_{S^\perp}^{(\ell-1)}
    =(I-S_{\ell-1}S_{\ell-1}^\dagger)\cdots
    (I-S_1S_1^\dagger).
\end{equation}
Each new block is then orthogonalized against $V$, $W$, and the previously generated blocks.

We also keep a history block $P$ from the previous iteration, with $P=\varnothing$ at the first iteration. Its update is described in Sec.~\ref{sec:gplhr_restart}. Before being added to the trial space, $P$ is orthogonalized against $V$, $W$, and $S_1,\ldots,S_m$. The full trial space is
\begin{equation}
    \mathcal Z=\operatorname{span}(Z),
    \qquad Z=[V,W,S_1,\ldots,S_m,P].
    \label{eq:trial_space}
\end{equation}
We next use harmonic Schur--Rayleigh--Ritz projection~\cite{Morgan1991, Morgan1998Harmonic} to obtain an improved approximation to the invariant subspace associated with eigenvalues near $\sigma$,
and update $V$, $M_A$, and $M_B$ accordingly. The extraction procedure
is described in the next subsection.

\subsubsection{Harmonic Schur--Rayleigh--Ritz procedure}
\label{sec:harmonic_extraction}
The next step is to extract a new $b$-dimensional
approximation to the invariant subspace associated with
eigenvalues near $\sigma$ from the expanded trial space.
Let $Z\in\mathbb C^{n\times d}$ be an orthonormal basis
of the trial space $\mathcal Z$, where $n$ is the dimension
of the Hilbert space under consideration and $d$ is the
dimension of the expanded trial space.
We write $V_{\mathrm{new}}=ZY$ with
$Y\in\mathbb C^{d\times b}$.
Assuming that $\sigma$ is not an eigenvalue of $H$,
we define the test basis
\begin{equation}
    U=\operatorname{orth}\!\left((H-\sigma I)Z\right).
    \label{eq:harmonic_test_basis}
\end{equation}
The harmonic projection requires the updated Schur residual to be
orthogonal to this test space, $U^\dagger R_{\mathrm{new}}=0$, giving
\begin{equation}
    (U^\dagger HZ)Y M_B^{\mathrm{new}}
    =(U^\dagger Z)Y M_A^{\mathrm{new}}.
    \label{eq:harmonic_projected_problem}
\end{equation}

We solve this small projected problem using an ordered
QZ decomposition of the $d\times d$ matrix pair
$(U^\dagger HZ,U^\dagger Z)$ \cite{Moler1973QZ}.
This produces unitary matrices
$\overline{Y}_L,\overline{Y}_R\in\mathbb C^{d\times d}$
of left and right Schur vectors, together with upper
triangular factors
$\overline{R}_A,\overline{R}_B\in\mathbb C^{d\times d}$.
The decomposition is ordered so that the first $b$
diagonal ratios
$(\overline{R}_A)_{jj}/(\overline{R}_B)_{jj}$
are closest to $\sigma$, with the corresponding Schur
vectors occupying the first $b$ columns of
$\overline{Y}_L$ and $\overline{Y}_R$.

We then retain
$Y_L=\overline{Y}_L(:,1:b)$ and
$Y_R=\overline{Y}_R(:,1:b)$.
The corresponding upper triangular factors are
$R_A=\overline{R}_A(1:b,1:b)$ and
$R_B=\overline{R}_B(1:b,1:b)$.
The selected Schur bases
$Y_L,Y_R\in\mathbb C^{d\times b}$ have orthonormal
columns and, together with
$R_A,R_B\in\mathbb C^{b\times b}$, satisfy
\begin{equation}
    (U^\dagger HZ)Y_R=Y_LR_A,
    \qquad
    (U^\dagger Z)Y_R=Y_LR_B.
    \label{eq:harmonic_schur_factors}
\end{equation}
The updated orthonormal basis is therefore
\begin{equation}
    V_{\mathrm{new}}=ZY_R.
    \label{eq:harmonic_basis_update}
\end{equation}

To compute $M_A^{\mathrm{new}}$ and $M_B^{\mathrm{new}}$, we set $Y=Y_R$ in Eq.~\eqref{eq:harmonic_projected_problem}
and substituting these relations gives
\begin{equation}
    Y_LR_AM_B^{\mathrm{new}}
    =Y_LR_BM_A^{\mathrm{new}}.
\end{equation}
Since $Y_L^\dagger Y_L=I$, multiplying both sides by
$Y_L^\dagger$ yields
\begin{equation}
    R_AM_B^{\mathrm{new}}=R_BM_A^{\mathrm{new}}.
    \label{eq:qfree_factor_condition}
\end{equation}
Then, following Ref.~\cite{VecharynskiEtAl2016}, writing $\alpha_j=(R_A)_{jj}$ and $\beta_j=(R_B)_{jj}$, we define diagonal matrices $D_1$ and $D_2$ by
\begin{equation}
    \bigl((D_1)_{jj},(D_2)_{jj}\bigr)
    =
    \begin{cases}
        \left(0,\dfrac{1}{\beta_j}\right),
        & |\alpha_j|<|\beta_j|,\\[6pt]
        \left(\dfrac{1-\beta_j}{\alpha_j},1\right),
        & \text{otherwise}.
    \end{cases}
\end{equation}
Then $\Gamma=R_AD_1+R_BD_2$ is upper triangular with unit
diagonal and hence invertible.
Solving $\Gamma J=R_A$ for the upper triangular
matrix $J$, we set
\begin{equation}
    M_A^{\mathrm{new}}=D_2J,
    \qquad
    M_B^{\mathrm{new}}=I-D_1J.
    \label{eq:qfree_factor_update}
\end{equation}
These factors satisfy Eq.~\eqref{eq:qfree_factor_condition},
ensuring the harmonic projection condition.
The corresponding eigenvalue estimates are given by
\begin{equation}
    \theta_j
    =\frac{(M_A^{\mathrm{new}})_{jj}}
           {(M_B^{\mathrm{new}})_{jj}}
    =\frac{\alpha_j}{\beta_j}.
\end{equation}

\subsubsection{Restart and eigenstate reconstruction.}
\label{sec:gplhr_restart}
After updating $V$, we form the new history block as
\begin{equation}
    P_{\mathrm{new}}
    =\operatorname{orth}\!\left[
    (I-\Pi_{V_{\mathrm{new}}})
    [0,W,S_1,\ldots,S_m,P]Y_R
    \right],
    \label{eq:history_update}
\end{equation}
where $\Pi_{V_{\mathrm{new}}}
=V_{\mathrm{new}}V_{\mathrm{new}}^\dagger$
is the orthogonal projector onto the updated subspace.
Here $Y_R$ is the coefficient matrix used to obtain
$V_{\mathrm{new}}=ZY_R$, and the zero block removes
the old-$V$ contribution. The remaining combination
is projected against $V_{\mathrm{new}}$ and then
orthonormalized.

The columns of the current basis $V$ span an approximation
to the target invariant subspace, but the basis vectors do not
in general correspond to individual eigenvectors. To recover a
right Ritz state, we form a linear combination
$\psi_i = Vd_i$ such that $H\psi_i \simeq \theta_i \psi_i$.

For nonsingular $M_B$, Eq.~\eqref{eq:qfree_schur} gives
$HV\simeq V(M_AM_B^{-1})$. Hence,
\begin{equation}
    H\psi_i=HVd_i\simeq V(M_AM_B^{-1})d_i.
\end{equation}
We then choose $d_i$ to satisfy the small eigenvalue problem
\begin{equation}
    (M_AM_B^{-1})d_i=\theta_i d_i.
\end{equation}
With $d_i = M_B c_i$, the problem becomes the equivalent generalized
eigenvalue problem
\begin{equation}
    M_Ac_i=\theta_iM_Bc_i,
    \label{eq:schur_to_ritz}
\end{equation}
which can be solved without explicitly forming $M_B^{-1}$. Once $c_i$ is obtained, the corresponding Ritz state is
\begin{equation}
    \psi_i=VM_Bc_i=Vd_i
    =\sum_{j=1}^{b}(d_i)_jv_j.
    \label{eq:gplhr-eigenstates}
\end{equation}
Each $\psi_i$ is then normalized to give the individual approximate
right eigenstate.

\subsection{Matrix product states}
\label{sec:fixed_D_arithmetic}
The GPLHR algorithm is a computationally efficient non-Hermitian eigensolver, but it still suffers from the curse of dimensionality. Therefore, we use the matrix product state (MPS) ansatz, which provides an efficient representation of quantum states and operators in one-dimensional many-body systems \cite{Schollwock2011,Orus2014}. Within the MPS ansatz, the wavefunction is written as

\begin{equation}
    \ket{\psi}=
    \sum_{s_1,\ldots,s_{n_s}}
    A_1^{s_1}A_2^{s_2}\cdots A_{n_s}^{s_{n_s}}
    \ket{s_1\cdots s_{n_s}},
    \label{eq:mps_representation}
\end{equation}
where $A_\ell^{s_\ell}$ has dimensions
$D_{\ell-1}\times D_\ell$ and $D_0=D_{n_s}=1$.
The Hamiltonian is represented as an MPO,
\begin{equation}
    H=\sum_{\boldsymbol s,\boldsymbol s'}
    B_1^{s_1s'_1}\cdots B_{n_s}^{s_{n_s}s'_{n_s}}
    \ket{\boldsymbol s}\bra{\boldsymbol s'},
    \label{eq:mpo_representation}
\end{equation}
with scalar boundary bonds.

The GPLHR iteration involves repeated Hamiltonian applications, linear combinations and orthogonalization of the trial vectors, as well as applications of the preconditioner, which approximates the shifted inverse $(H-\sigma I)^{-1}$. When the vectors are represented as MPSs, these operations tend to increase their bond dimensions. In particular, an exact application of the Hamiltonian in MPO form multiplies the MPS bond dimension by the corresponding MPO bond dimension, whereas linear combinations of MPSs can increase the bond dimension through direct sums of their virtual spaces. Similar growth can arise from the application of the preconditioner, depending on its tensor-network realization. Without truncation, the bond dimensions would therefore grow rapidly over successive GPLHR iterations, making the calculation increasingly expensive. Compression is consequently necessary throughout the iteration to restrict the bond dimension to a prescribed maximum value $D$.

To perform this compression, we use successive randomized
compression (SRC)~\cite{CamanoEtAl2026}, which approximates
MPO--MPS products with a maximum bond dimension $D$
without explicitly forming the full uncompressed product MPS. Its basic operation is a randomized
approximation to the column space of a matrix $K$:
\begin{equation}
    Y=K\Omega,\qquad Q=\operatorname{orth}(Y),
    \qquad K\approx QQ^\dagger K,
    \label{eq:randomized_qb}
\end{equation}
where $\Omega$ is a random test matrix.
Starting from the rightmost site, SRC applies this
approximation successively to matrix unfoldings of
the MPO--MPS product. At each step, a QR factorization
provides a local orthonormal basis, which defines
the output tensor and is contracted into the
remaining network. The sweep produces an MPS
in right-canonical form.
Structured random sketches built from local Gaussian
matrices allow these operations to be evaluated
through tensor contractions, while intermediate
contractions are reused to reduce computational cost.

\subsection{GPLHR with matrix product states}
\label{subsec:mps-gplhr}

We now combine the GPLHR framework introduced in
Sec.~\ref{sec:gplhr_algorithm} with the MPS representation.
The resulting MPS-GPLHR algorithm follows the same
subspace expansion, harmonic extraction, and restart
steps, with each vector represented as an MPS and the
Hamiltonian represented as an MPO. The full procedure is
summarized in Algorithm~\ref{alg:gplhr}.

The iteration starts from a block of linearly independent
MPSs, $V^{(0)}=[v_1^{(0)},\ldots,v_b^{(0)}]$, and a
target shift $\sigma$ specifying the spectral region
of interest.
The initial guesses play an important role in MPS-GPLHR.
Here, the initial guesses refer to the MPSs in
$V^{(0)}$, whose span forms the initial approximation
to the target invariant subspace.
Their overlap with the targeted spectral components
influences convergence.
Suitable choices include product states and random
MPSs with small bond dimensions.

Before the iteration, the initial block is compressed
to the bond cap $D$, normalized, and orthogonalized.
An initial harmonic extraction, using the compressed
Hamiltonian actions and the target shift $\sigma$,
determines the starting subspace and triangular factors
$M_A,M_B$.
These factors are used to form the first Schur residual.
The shift is chosen near the eigenvalues
of interest. The specific initial states and target shifts used
for each benchmark are given in
Sec.~\ref{sec:models}.

The essential step in GPLHR is to apply an MPO to an MPS, and this increases
the MPS bond dimension. Therefore, we define $\mathcal C_D$ as the compression operator that reduces the resulting MPS to a maximum bond dimension $D$. Inner products and projected matrix elements are evaluated
by tensor contractions.

The construction of the trial space requires each new search direction
to be orthogonalized against the existing basis. For a basis block $B$, the routine
$\operatorname{Proj}_D(x;B)$ removes the component of $x$ along $B$
and then compresses the result:
\begin{equation}
    x \leftarrow
    \mathcal C_D\!\left[
        x-B(B^\dagger B)^+B^\dagger x
    \right].
\end{equation}
This form allows for basis vectors that are not exactly orthonormal
after MPS compression. Here, $+$ denotes the pseudoinverse, with
numerically null modes discarded. Since compression can introduce
overlap with the existing basis, the projection--compression step is
applied twice. The final state is normalized if its norm is above a
prescribed threshold; otherwise, it is discarded.

The Hamiltonian action on the current subspace is stored as
$F_V=\mathcal C_D(HV)$.
We form the Schur residual as $F_V M_B - V M_A$. Since these MPS linear combinations can increase
the bond dimension, we compress the residual to a maximum bond
dimension $D$ before using it to generate correction directions.
The projections and orthonormalizations in
Sec.~\ref{sec:gplhr_residual} are implemented through
successive applications of $\operatorname{Proj}_D$.
This produces the correction block $W$, any additional
blocks $S_1,\ldots,S_m$, and the projected history block
$P$. The trial space is constructed as in
Eq.~\eqref{eq:trial_space}. By default, we set $m=0$ to avoid the extra MPO applications
and compression steps associated with the expansion blocks.
If the iteration stalls, we optionally add one expansion block $S_1$.

The shifted preconditioner $T_\sigma$ is implemented
by a short MPS-GMRES solve of
$(H-\sigma I)u=q$.
The resulting MPS $u$ approximates $T_\sigma q$,
with compression applied during the solve to control
the bond dimension.
An iteration cap $m_{\mathrm{in}}$ limits the
computational cost. GMRES approximates the action of $T_\sigma$ on the
projected residual. The resulting MPS is then projected
against $V$ to obtain the action of $T_L$.

For the harmonic Schur--Rayleigh--Ritz procedure, let
\begin{equation}
    A_Z=\mathcal C_D(HZ),
    \qquad
    Q_Z=A_Z-\sigma Z,
    \qquad
    G=Z^\dagger Z.
\end{equation}
The block $Q_Z$ approximates the shifted trial block used to construct the test basis in
Eq.~\eqref{eq:harmonic_test_basis}. Since the harmonic projection condition depends only
on the test space, $Q_Z$ need not be explicitly orthonormalized. We then form the small projected matrix pair as in Eq.~\eqref{eq:harmonic_projected_problem}
\begin{equation}
    (\mathsf A,\mathsf B)
    =
    (Q_Z^\dagger A_Z,\;Q_Z^\dagger Z).
\end{equation}
We next apply an ordered QZ decomposition to the small
projected matrix pair. Since the MPS trial basis $Z$
is not necessarily orthonormal, a transformation $X$
is first constructed such that
\begin{equation}
    X^\dagger GX=I,
\end{equation}
so that $ZX$ forms an orthonormal basis of the retained
trial space. In these coordinates, the projected
matrix pair becomes
\begin{equation}
    (X^\dagger\mathsf A X,\;X^\dagger\mathsf B X).
\end{equation}
Its ordered QZ decomposition yields the selected
coefficient matrices $Y_L,Y_R$ and upper triangular
matrices $R_A,R_B$, corresponding to the $b$ finite
generalized eigenvalues closest to $\sigma$.
The coefficients are then mapped back to the original
trial basis through
\begin{equation}
    Y=XY_R.
\end{equation}
The factors $M_A^{\mathrm{new}}$ and $M_B^{\mathrm{new}}$
are constructed from $R_A,R_B$ using
Eq.~\eqref{eq:qfree_factor_update}. The updated MPS basis is then formed as
\begin{equation}
    V_{\mathrm{new}}=\mathcal C_D(ZY).
\end{equation}

During the iteration, the Schur residual computed from
the stored compressed Hamiltonian actions is used
to determine convergence.
The iteration also stops when the prescribed stagnation
criterion is met or the iteration limit is reached.
Upon termination, the right eigenstates are recovered
using Eq.~\eqref{eq:gplhr-eigenstates}, compressed,
and normalized.
Their final eigenpair residuals are evaluated separately
to assess accuracy, without changing the solver's
convergence flag.

\begin{algorithm}[htbp]
\caption{MPS--GPLHR}
\label{alg:gplhr}
\begin{algorithmic}[1]
\Require MPO $H$, $b$ initial MPSs $V$, target $\sigma$,
         bond cap $D$, tolerance $\varepsilon$,
         maximum iterations $K_{\max}$
\Ensure Approximate eigenpairs
        $\{(E_i,|\psi_i\rangle)\}_{i=1}^{b}$
        and convergence status

\State Compress and orthogonalize $V$;
       $P\gets\varnothing$
\State $F_V\gets\mathcal C_D(HV)$
\State $(Y,M_A,M_B,G)\gets
       \operatorname{Harmonic}(V,F_V)$
\State $V\gets\mathcal C_D(VY)$
\State $F_V\gets\mathcal C_D(HV)$
\State $\mathrm{converged}\gets\mathrm{false}$

\For{$k=1,\ldots,K_{\max}$}
    \State Evaluate by tensor contractions:
           $\rho_{\mathrm S}\gets
           \max_j\|(F_VM_B-VM_A)_j\|_2$
    \If{$\rho_{\mathrm S}\leq\varepsilon$}
        \State $\mathrm{converged}\gets\mathrm{true}$
        \State \textbf{break}
    \EndIf

    \State $R\gets\mathcal C_D(F_VM_B-VM_A)$
    \State $W\gets[\,]$
    \For{$j=1,\ldots,b$}
        \State $q_j\gets\operatorname{Proj}_D(R_j;V)$
        \If{$q_j$ is retained}
            \State $u_j\gets T_\sigma(q_j)$
            \State $w_j\gets
                   \operatorname{Proj}_D(u_j;[V,W])$
            \State Append $w_j$ to $W$ if retained
        \EndIf
    \EndFor

    \State Project and orthonormalize $P$ against
           $[V,W]$ using $\operatorname{Proj}_D$,
           discarding dependent directions
    \State $Z\gets[V,W,P]$
    \State $F_Z\gets
           [F_V,\mathcal C_D(HW),\mathcal C_D(HP)]$
    \State $(Y,M_A,M_B,G)\gets
           \operatorname{Harmonic}(Z,F_Z)$

    \State $Y_P\gets Y$ with its first $b$ rows set to zero
    \State $Y_P\gets Y_P-Y(Y^\dagger GY_P)$
    \State Orthonormalize $Y_P$ in the metric $G$,
           discarding dependent directions

    \State $V\gets\mathcal C_D(ZY)$
    \State $P\gets\mathcal C_D(ZY_P)$
    \State $F_V\gets\mathcal C_D(HV)$
\EndFor

\State Solve $M_Ac_i=\theta_iM_Bc_i$
       for $i=1,\ldots,b$
\For{$i=1,\ldots,b$}
    \State $\widehat\psi_i\gets\mathcal C_D(VM_Bc_i)$
    \State $\psi_i\gets
           \widehat\psi_i/\|\widehat\psi_i\|_2$
    \State $f_i\gets\mathcal C_D(H\psi_i)$,
           $E_i\gets\langle\psi_i|f_i\rangle$
    \State Evaluate
           $\rho_i\gets\|H\psi_i-E_i\psi_i\|_2$
           using contractions with the original MPO $H$
\EndFor
\State \Return
       $\{(E_i,|\psi_i\rangle)\}_{i=1}^{b}$,
       $\mathrm{converged}$
\end{algorithmic}
\end{algorithm}

\section{NUMERICAL EXPERIMENTS}
\label{sec:models}
We benchmark the proposed MPS-GPLHR scheme on representative non-Hermitian Hamiltonians by monitoring both the convergence of the targeted eigenpairs and the GPLHR iterations required to obtain them.  To estimate the error of the computed eigenstates, we use the squared eigenresidual, analogous to the energy variance commonly employed for Hermitian eigenvalue problems:
\begin{equation}
    \lVert r \rVert^2
    =
    \frac{
        \lVert (H-E)\lvert\psi\rangle \rVert^2
    }{
        \langle\psi\vert\psi\rangle
    }.
\end{equation}
In this work, the tensor network backend uses Renormalizer and NumPy/SciPy, with CuPy for GPU tensor kernels
\cite{HarrisEtAl2020,VirtanenEtAl2020,OkutaEtAl2017}.
All MPOs used in this paper are constructed automatically following graph-based operator assembly~\cite{RenEtAl2020,CakirEtAl2025}.

\subsection{Hubbard chains}
\paragraph{Transcorrelated Hubbard model}\label{hb} We first benchmark the MPS-GPLHR method on the transcorrelated Hubbard model with open boundary conditions \cite{Hubbard1963,LiebWu1968}. The Hubbard model provides a standard and well-characterized test bed for strong electronic correlation, while its transcorrelated form offers a particularly useful benchmark for non-Hermitian eigensolvers. The transcorrelated Hamiltonian is obtained through a non-unitary similarity transformation of the original Hermitian Hamiltonian \cite{BoysHandy1969,BaiardiReiher2020}. Consequently, although the transformed Hamiltonian is non-Hermitian, it is isospectral to the original Hamiltonian. All eigenvalues of transcorrelated Hubbard models tested here remain unchanged and real. At the same time, the transformation incorporates part of the many-body correlation into the Hamiltonian, which can lead to a more compact representation of the corresponding eigenstates.  We consider the one-dimensional fermonic Hubbard model with open boundary conditions,
\begin{align}
 H={}&-\sum_{j=0}^{L-2}\sum_s
 \left(t_Rc^\dagger_{j+1,s}c_{j,s}+t_Lc^\dagger_{j,s}c_{j+1,s}\right)
 \nonumber\\
 &+U\sum_{j=0}^{L-1}n_{j\uparrow}n_{j\downarrow}
 +\sum_{j=0}^{L-1}v_j(n_{j\uparrow}+n_{j\downarrow}),
 \label{eq:hubbard}
\end{align}
where $s\in\{\uparrow,\downarrow\}$ labels the spin.
We work at half filling with
$N_\uparrow=N_\downarrow=L/2$.
Energies are expressed in units of the hopping amplitude,
and we set $U=4$.

\begin{figure}[htbp]
    \centering

    \begin{subfigure}[t]{0.48\linewidth}
        \centering
        \includegraphics[width=\linewidth]{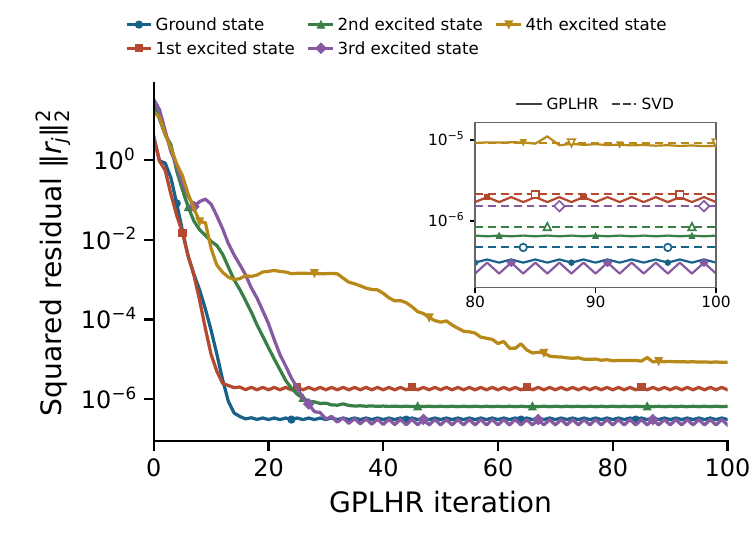}
        \caption{Convergence for the transcorrelated Hubbard model.}
        \label{fig:real-convergence}
    \end{subfigure}
    \hfill
    \begin{subfigure}[t]{0.48\linewidth}
        \centering
        \includegraphics[width=\linewidth]{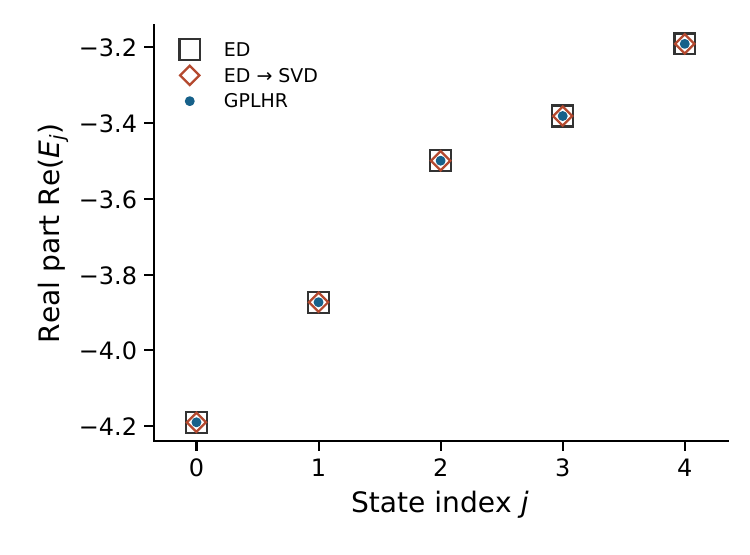}
        \caption{Spectrum comparison for the transcorrelated Hubbard model.}
        \label{fig:real-spectrum}
    \end{subfigure}

    \vspace{0.8em}

    \begin{subfigure}[t]{0.48\linewidth}
        \centering
        \includegraphics[width=\linewidth]{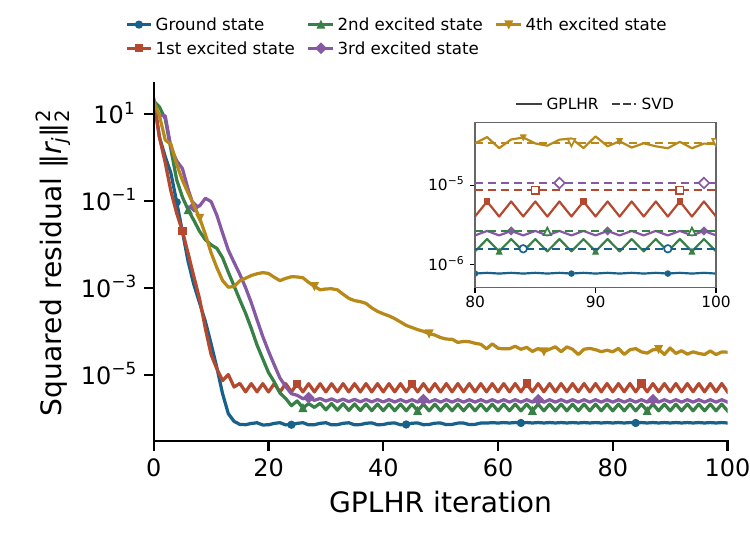}
        \caption{Convergence for the complex Hubbard model.}
        \label{fig:complex-convergence}
    \end{subfigure}
    \hfill
    \begin{subfigure}[t]{0.48\linewidth}
        \centering
        \includegraphics[width=\linewidth]{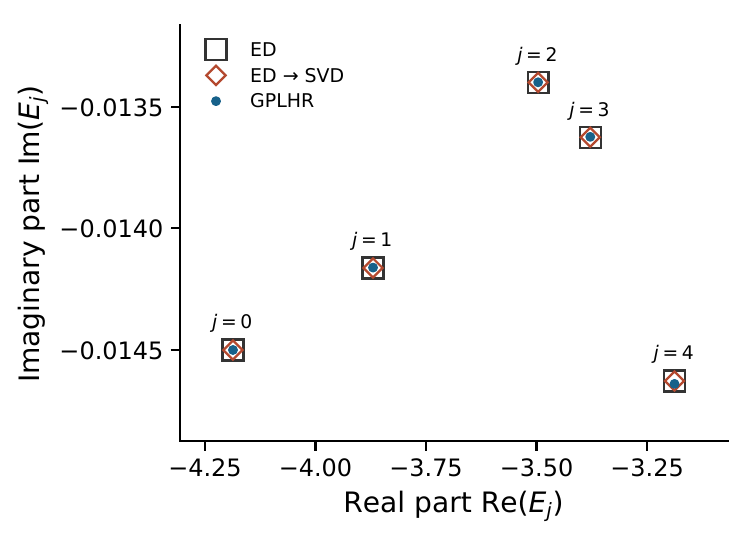}
        \caption{Spectrum comparison for the complex Hubbard model.}
        \label{fig:complex-spectrum}
    \end{subfigure}

    \caption{
    Convergence and spectra of the first five right eigenstates of an open,
    half-filled $L=8$ Hubbard chain with $U=4$ and bond dimension $D=64$.
    Solid lines show the GPLHR results, while dashed lines show the corresponding
    SVD-compressed exact-state references.
    (a) Squared residuals for the transcorrelated Hubbard model with
    $t_L=t_R=\sqrt{0.99}$, $g=0.7$, and $\sigma=-5$.
    (b) Spectrum comparison for the same model.
    (c) Squared residuals for the Hubbard model with the
    complex potential of Eq.~\eqref{eq:complex_potential}, with
    $t_R=1.10$, $t_L=0.90$, and $\sigma=-5-0.014i$.
    (d) Spectrum comparison for the same model.
    Panels (b) and (d) compare the eigenvalues from ED, SVD-compressed exact states,
    and MPS-GPLHR.
    }
    \label{fig:spectrum-and-convergence}
\end{figure}

For the transcorrelated benchmark, we start from a Hermitian Hubbard Hamiltonian with symmetric hopping,
$t_L=t_R=\sqrt{0.99}$, and a real quasiperiodic onsite potential,
\begin{equation}
    v_j=0.15\cos[2\pi\phi(j+1)+0.37],\qquad
    \phi=(\sqrt5-1)/2.
    \label{eq:qp}
\end{equation}
The symmetric hopping and real onsite potential ensure that the reference Hamiltonian $H$ is Hermitian. We then construct a controlled non-Hermitian problem by applying the non-unitary transcorrelated similarity transformation
\begin{equation}
    H_{\tc}=e^{-\tau}He^\tau,\qquad
    \tau=g\sum_j n_{j\uparrow}n_{j\downarrow},\qquad g=0.7.
    \label{eq:tc}
\end{equation}
Because this transformation is a similarity transformation, $H_{\tc}$ has the same spectrum as the original Hermitian Hamiltonian, while its right eigenstates are modified and the transformed Hamiltonian is generally non-Hermitian. This provides a convenient benchmark for testing the convergence and accuracy of the non-Hermitian eigensolver.

For this benchmark, the initial block consists
of the two N\'eel product states
$|\uparrow\downarrow\uparrow\downarrow\cdots\rangle$
and $|\downarrow\uparrow\downarrow\uparrow\cdots\rangle$,
together with three random MPSs initialized with a
maximum bond dimension of four.
All initial states belong to the sector
$N_\uparrow=N_\downarrow=L/2$ and are normalized
and orthogonalized before the iteration.
We set the target shift to $\sigma=-5$.

Fig.~\ref{fig:spectrum-and-convergence}(a) presents the error of 
the first five right eigenstates obtained by GPLHR for a half-filled $L=8$ transcorrelated Hubbard chain with bond dimension $D=64$, using the results from exact diagonalization (ED) as reference.
To test the accuracy of GPLHR, we compare the errors of our method with the so-called SVD-compressed states, i.e., the resulting states obtained by compressing an exact eigenstate (in MPS form) with bond dimension $D$. For the transcorrelated Hubbard model considered here, the GPLHR residuals converge to approximately the same level as those of the SVD-compressed reference states. Fig.~\ref{fig:spectrum-and-convergence}(b) compares the corresponding real eigenvalues, which agree closely with the ED and SVD results. These results show that MPS-GPLHR achieves accuracy comparable to SVD compression at the same bond dimension.

\begin{figure}[htbp]
    \centering

    \begin{subfigure}[t]{0.8\linewidth}
        \centering
        \includegraphics[width=\linewidth]{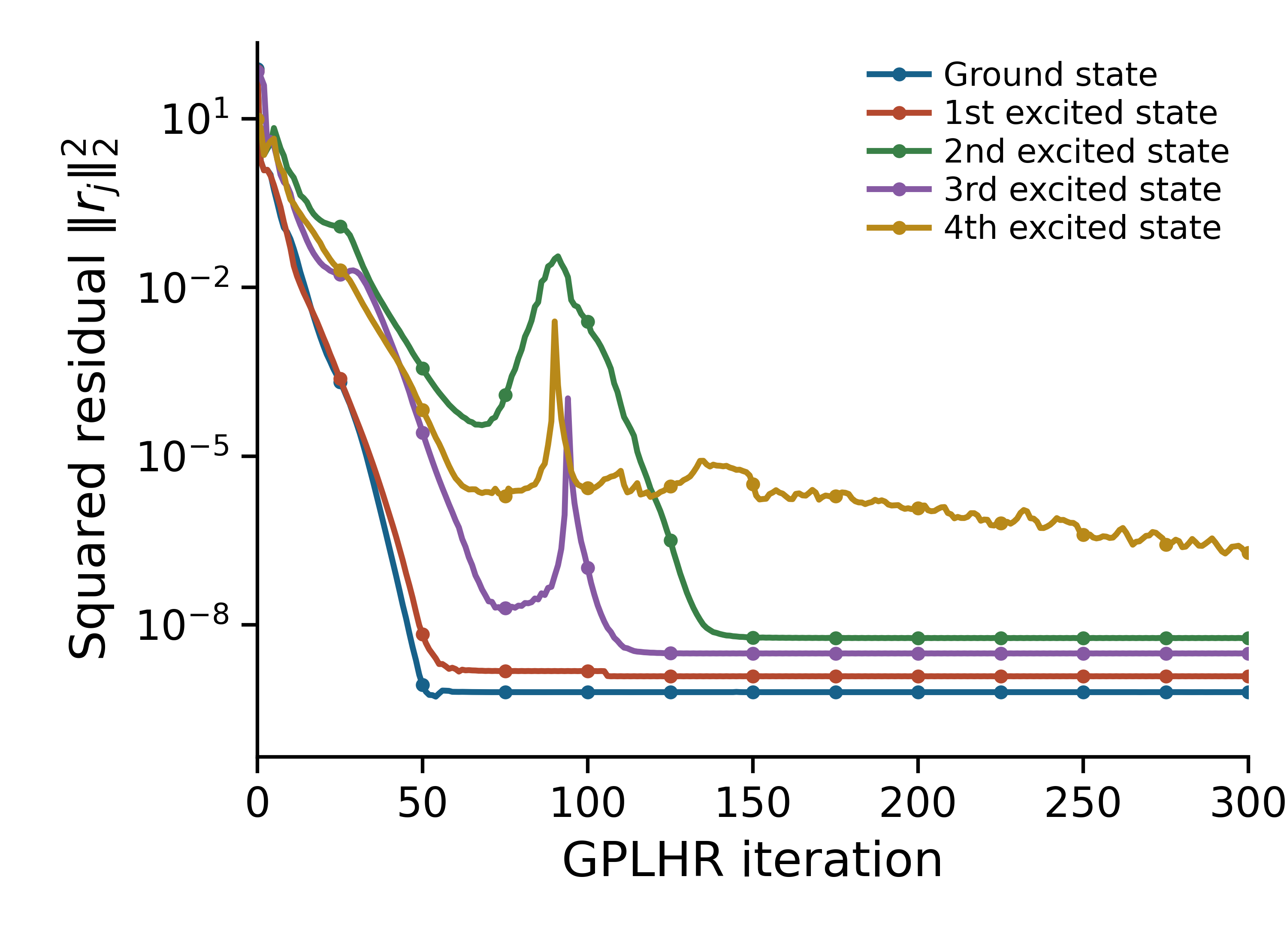}
        \caption{Residual convergence for $L=20$.}
        \label{fig:L20}
    \end{subfigure}

    \vspace{0.5em}

    \begin{subfigure}[t]{0.8\linewidth}
        \centering
        \includegraphics[width=\linewidth]{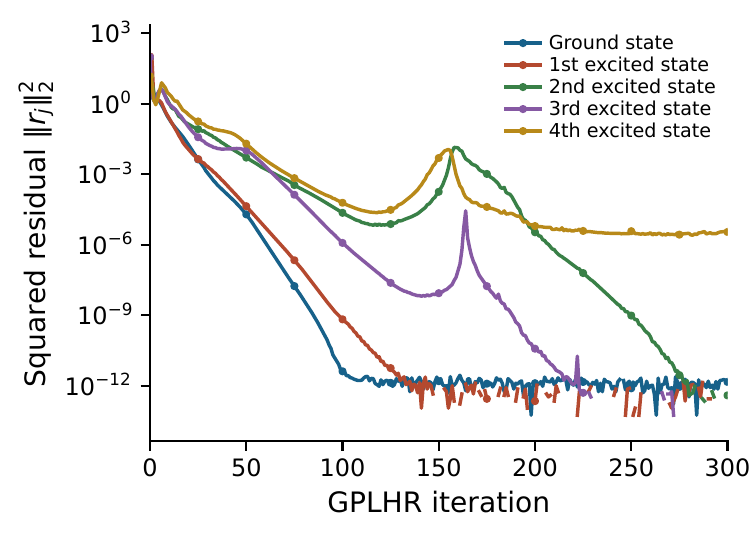}
        \caption{Residual convergence for $L=30$.}
        \label{fig:L30}
    \end{subfigure}

    \caption{
        Convergence of the first five right eigenstates of the half-filled
        transcorrelated Hubbard chain.
        The squared residuals are shown as a function of the GPLHR iteration.
        (a) $L=20$ with $D=256$ and $\sigma=-12$.
        (b) $L=30$ with $D=512$ and $\sigma=-18$.
    }
    \label{fig:L2030}
\end{figure}

Fig.~\ref{fig:L2030} shows the convergence of the first five right eigenstates for two larger systems: $L=20$ with $D=256$ and $\sigma=-12$, and $L=30$ with $D=512$ and $\sigma=-18$. Both calculations extend to 300 iterations on an A100 GPU. For $L=20$, four trajectories settle near the $10^{-9}$ scale in squared residual, while the fifth remains near $10^{-7}$. For $L=30$, four
reach the numerical floor displayed near $10^{-12}$ and the remaining trajectory is of order $10^{-6}$.

\begin{figure}[htbp]
    \centering

    \begin{subfigure}[t]{0.8\linewidth}
        \centering
        \includegraphics[width=\linewidth]{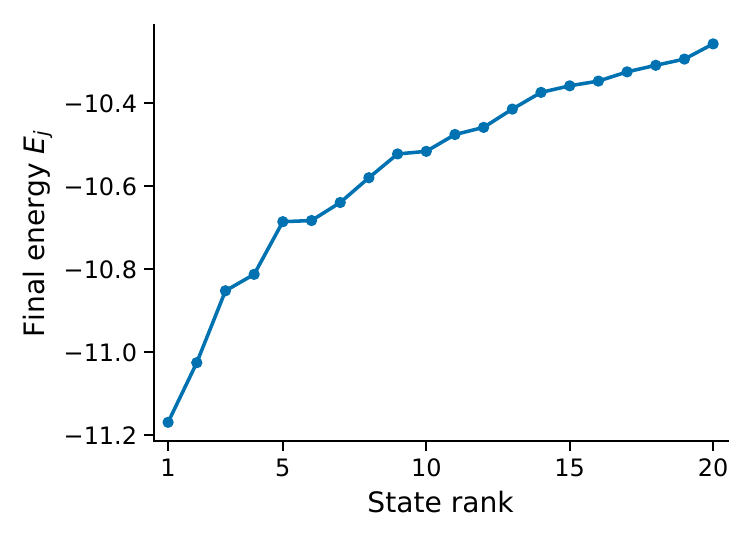}
        \caption{Energies of the converged states.}
        \label{fig:L20_energy}
    \end{subfigure}

    \vspace{0.5em}

    \begin{subfigure}[t]{0.8\linewidth}
        \centering
        \includegraphics[width=\linewidth]{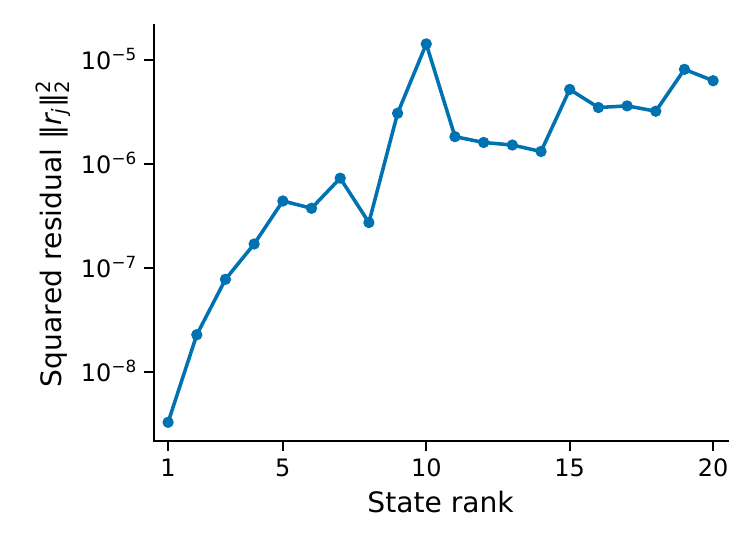}
        \caption{Squared residuals of the converged states.}
        \label{fig:L20_residual}
    \end{subfigure}

    \caption{
        Final results for the lowest 20 right eigenstates of an open,
        half-filled $L=20$ transcorrelated Hubbard chain with bond dimension
        $D=256$.
        (a) Energies of the converged states, ordered by increasing energy.
        (b) Corresponding squared residuals.
    }
    \label{fig:20eigs}
\end{figure}

We further illustrate GPLHR's ability to compute higher-lying
eigenstates. Fig.~\ref{fig:20eigs} collects twenty low-lying
states for $L=20$ and $D=256$, obtained from five overlapping
spectral windows with target shifts
$\sigma=-12.000000$, $-10.692935$, $-10.518163$,
$-10.401280$, and $-10.299815$, respectively. For the larger retained blocks used in these calculations,
additional product states are generated by exchanging an
up-spin and a down-spin occupation in the N\'eel configurations,
and random MPSs complete the block.
Each window is initialized independently and evolved for
150 GPLHR iterations.
The final squared residuals shown in the figure range
approximately from $10^{-8}$ to $10^{-5}$.
In practice, the shift $\sigma$ can be updated sequentially
using information from the eigenvalues obtained in the
preceding window, thereby enabling the targeted exploration
of progressively higher-lying states.

\paragraph{Complex Hubbard model}

We further benchmark the ability of GPLHR to handle complex spectra by considering a Hubbard model with a genuinely complex spectrum. Here, we set $\tau=0$, $t_R=1.10$, $t_L=0.90$, and introduce the complex on-site potential
\begin{align}
 v_j={}&0.15\cos[2\pi\phi(j+1)+0.37]
 \nonumber\\
 &+0.12i\sin[2\pi\phi(j+1)+0.83].
 \label{eq:complex_potential}
\end{align}
The imaginary component of the on-site potential introduces
spatially dependent gain and loss.
Fig.~\ref{fig:spectrum-and-convergence}(c) shows the
squared eigenpair residuals for the first five right
eigenstates of a half-filled $L=8$ complex Hubbard chain.
We use the same GPLHR settings as in the transcorrelated
benchmark, except for the target shift
$\sigma=-5-0.014i$.
Several residuals approach levels comparable to those
of the SVD-compressed exact-state references. Fig.~\ref{fig:spectrum-and-convergence}(d) compares the corresponding eigenvalues in the complex plane. The MPS-GPLHR results agree closely with the ED and SVD results. These results demonstrate that the MPS-based GPLHR approach can accurately target eigenstates even when the spectrum is genuinely complex.

\subsection{Interacting non-Hermitian SSH chain}
To further assess GPLHR for interacting lattice systems, we consider a non-Hermitian extension of the Su-Schrieffer-Heeger (SSH) chain~\cite{SuEtAl1979,ZhongEtAl2025}. Let $a_\ell=c_{2\ell-1}$ and $b_\ell=c_{2\ell}$ denote
the fermionic annihilation operators on the two sites
of unit cell $\ell$. With open boundary conditions and
$u=0$, the Hamiltonian is
\begin{align}
H_{\mathrm{SSH}}
={}&\sum_{\ell=1}^{N_c}
\left(t_R b_\ell^\dagger a_\ell
+t_L a_\ell^\dagger b_\ell\right)
\nonumber\\
&+t_2\sum_{\ell=1}^{N_c-1}
\left(a_{\ell+1}^\dagger b_\ell
+b_\ell^\dagger a_{\ell+1}\right)
\nonumber\\
&+V\sum_{j=1}^{L-1}n_jn_{j+1},
\label{eq:ssh_spinless}
\end{align}
where $L=2N_c$, $n_j=c_j^\dagger c_j$, $t_R=t_1+\gamma$, and $t_L=t_1-\gamma$. We first benchmark its scalability in the real-spectrum regime of the interacting SSH model. We set $t_2=1$, $t_1=0.7$, and $\gamma=0.1$, corresponding to asymmetric intracell hoppings $t_R=t_1+\gamma=0.8$ and $t_L=t_1-\gamma=0.6$, together with $V=5$ and $u=0$. We consider an open chain with $N_c=64$ unit cells, corresponding to $L=128$ spinless-fermion sites, at half filling with $N_{\rm p}=64$ particles.  Although $t_R\ne t_L$ makes the Hamiltonian non-Hermitian, its eigenvalues are real in this case. For open boundaries and $u=0$, a nonunitary diagonal similarity transformation maps the hopping
terms to Hermitian ones with intracell amplitude
$\sqrt{t_Lt_R}$; the density interaction is unchanged by this
transformation. Hence, the full interacting Hamiltonian is similar
to a Hermitian Hamiltonian and has a real spectrum.

\begin{figure}[t]
  \centering
  \includegraphics[width=\columnwidth]
    {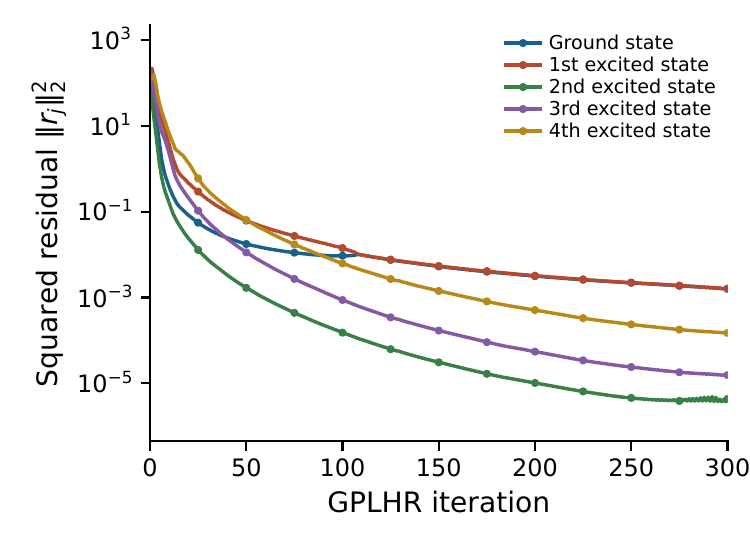}
  \caption{GPLHR convergence for the first five right eigenpairs
  of the half-filled, 128-site interacting SSH chain with $u=0$.}
  \label{fig:ssh128-residual}
\end{figure}

Fig.~\ref{fig:ssh128-residual} shows the evolution of the residuals of the first five eigenpairs over 300 GPLHR iterations. For this 128-site benchmark, the initial block contains
the two alternating occupation product states
$|1010\cdots\rangle$ and $|0101\cdots\rangle$,
together with three random MPSs initialized with a
maximum bond dimension of four.
All five states belong to the half-filled sector
$N_{\rm p}=L/2=64$ and are normalized and orthogonalized
before the iteration. We set the target shift to $\sigma=-20$.

The results demonstrate that MPS-GPLHR can effectively handle large-scale systems far beyond the reach of conventional methods based on explicit matrix representations. We further consider the complex interacting spinless SSH
chain with open boundaries and $L=20$ sites.
Each site has two local Fock states, $|0\rangle$ and
$|1\rangle$, and we work at half filling with
$N_{\rm p}=L/2=10$.
The hopping parameters are $t_R=0.8$, $t_L=0.6$,
and $t_2=1$.
The model includes a nearest-neighbor density interaction
$V\sum_{j=1}^{L-1}n_jn_{j+1}$ and a complex staggered
potential with $u=0.2$, where $n_j=c_j^\dagger c_j$.

Let $a_\ell=c_{2\ell-1}$ and $b_\ell=c_{2\ell}$ denote the two
spinless-fermion sites in cell $\ell=1,\ldots,N_c$, with $L=2N_c$ physical sites. For open boundaries, we use
\begin{align}
H_{\rm SSH}={}&\sum_{\ell=1}^{N_c}
(t_R b_\ell^\dagger a_\ell+t_L a_\ell^\dagger b_\ell)
\nonumber\\
&+t_2\sum_{\ell=1}^{N_c-1}
(a_{\ell+1}^\dagger b_\ell+b_\ell^\dagger a_{\ell+1})
\nonumber\\
&+V\sum_{j=1}^{L-1}n_jn_{j+1}
+(1-i)u\sum_{\ell=1}^{N_c}
(n_{a_\ell}-n_{b_\ell}).
\label{eq:ssh}
\end{align}

\begin{figure}[t]
  \centering
  \includegraphics[width=\columnwidth]
    {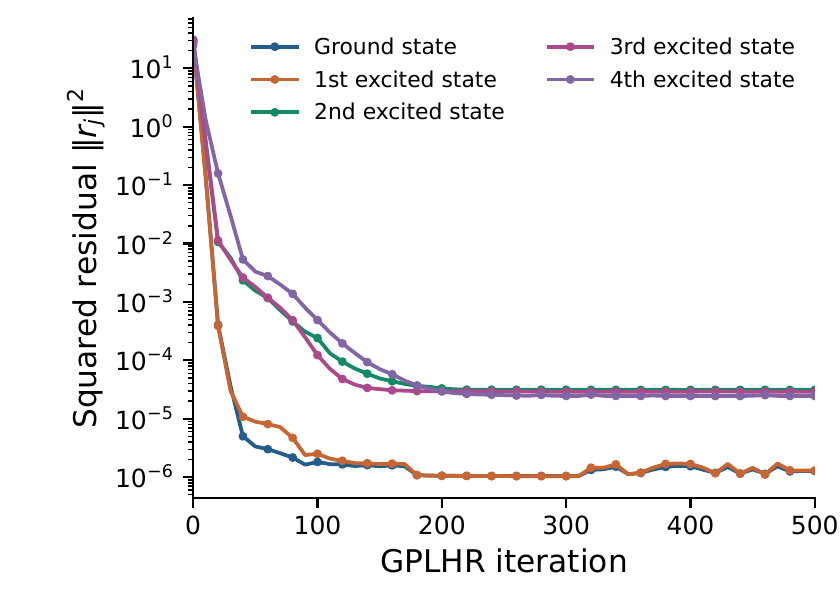}
  \caption{Squared residuals of five targeted right-eigenpair
  approximations for the 20-site interacting spinless non-Hermitian SSH chain.
  GPLHR uses MPS bond cap $D=200$ and complex target
  $\sigma=20+0.15i$. Residuals are evaluated with the
  original Hamiltonian every ten iterations.}
  \label{fig:ssh20-spinful-residual}
\end{figure}

Figure~\ref{fig:ssh20-spinful-residual} shows the convergence of the five target eigenstates over 500 GPLHR updates. For this five-state benchmark, the initial block consists
of the two alternating occupation product states
$|1010\cdots\rangle$ and $|0101\cdots\rangle$,
together with three random MPSs initialized with a
maximum bond dimension of four.
All states belong to the half-filled sector
$N_{\rm p}=10$ and are normalized and orthogonalized
before the iteration.
We set the target shift to $\sigma=20+0.15i$, guided by
low-energy estimates from an independent six-site calculation.

In contrast to the preceding real-spectrum benchmark, this spinless model provides a genuinely non-Hermitian test in which the target eigenvalues are complex. The steady reduction of the residuals demonstrates that MPS-GPLHR remains effective in the presence of both strong many-body interactions and complex spectral structure, and can reliably resolve multiple complex eigenpairs in a Hilbert space that is already prohibitively large for explicit matrix-based eigensolvers.

\subsection{Periodic quadratic Liouvillian}
\label{sec:quadratic-liouvillian}

Last, we benchmark MPS-GPLHR on a Liouvillian system of $L$ spinless fermionic sites, whose density operator evolves according to
\begin{equation}
  \frac{d\rho}{dt}=\mathcal{L}(\rho)
  =-i[H,\rho]+\sum_{j=1}^{L}\left(
    F_j\rho F_j^\dagger-\frac{1}{2}\{F_j^\dagger F_j,\rho\}
  \right).
\end{equation}
Writing $d_j=c_j-e^{-i\phi}c_{j+1}$, the Hamiltonian and jump operators are
\begin{equation}
  H=J\sum_{j=1}^{L}d_j^\dagger d_j,
  \qquad F_j=\sqrt{\kappa}\,d_j,
  \qquad c_{L+1}=c_1.
\end{equation}
We set $\kappa=1$, $J=0.25$, and $\phi=\pi/L$.
The quadratic Hamiltonian and linear loss operators make this a quasi-free model with an analytically accessible spectrum; the general spectral framework for such
Lindbladians is discussed in Ref.~\cite{Barthel2022Solving}.

To apply an eigensolver, we express the master equation as a matrix acting on a vectorized density operator.
We use the convention
\begin{equation}
  |\rho\rangle\!\rangle
  =\sum_{n,m=0}^{D-1}\rho_{nm}|n\rangle\otimes|m\rangle,
  \qquad
  |A\rho B\rangle\!\rangle
  =(A\otimes B^{\mathsf T})|\rho\rangle\!\rangle.
\end{equation}
Thus $\rho\in\mathbb{C}^{D\times D}$ becomes a vector of length
$N_H=4^L$, and $\mathcal{L}$ becomes the non-Hermitian matrix
\begin{equation}
\begin{aligned}
  \mathbb{L}={}&-i(H\otimes I_D-I_D\otimes H^{\mathsf T})\\
  &+\sum_j\left[
    F_j\otimes F_j^*
    -\frac{1}{2}F_j^\dagger F_j\otimes I_D
    -\frac{1}{2}I_D\otimes(F_j^\dagger F_j)^{\mathsf T}
  \right],
\end{aligned}
\label{eq:vectorized-liouvillian}
\end{equation}
of dimension $4^L\times4^L$, with
$\partial_t|\rho\rangle\!\rangle=\mathbb{L}|\rho\rangle\!\rangle$.
Here $\mathsf T$ and $*$ denote transpose and entrywise complex conjugation.

\begin{figure*}[t]
  \centering
  \begin{minipage}[t]{0.49\textwidth}
    \centering
    \makebox[\linewidth][l]{\textbf{(a)}}\par
    \includegraphics[width=\linewidth]{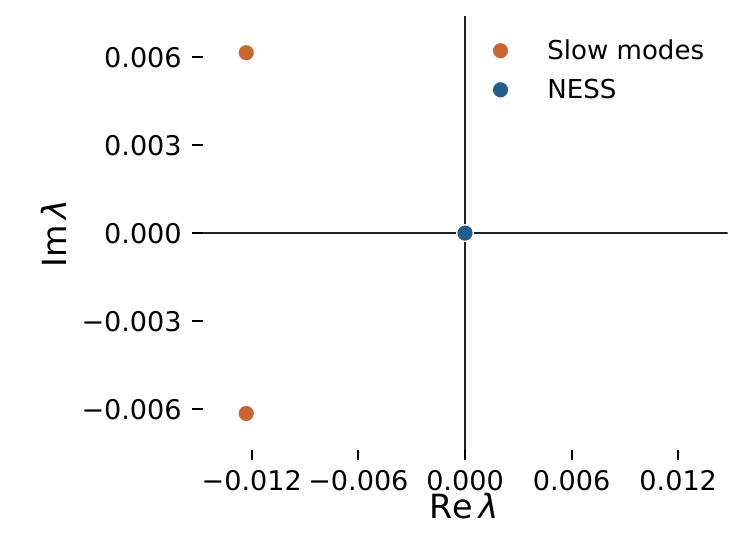}
  \end{minipage}\hfill
  \begin{minipage}[t]{0.49\textwidth}
    \centering
    \makebox[\linewidth][l]{\textbf{(b)}}\par
    \includegraphics[width=\linewidth]{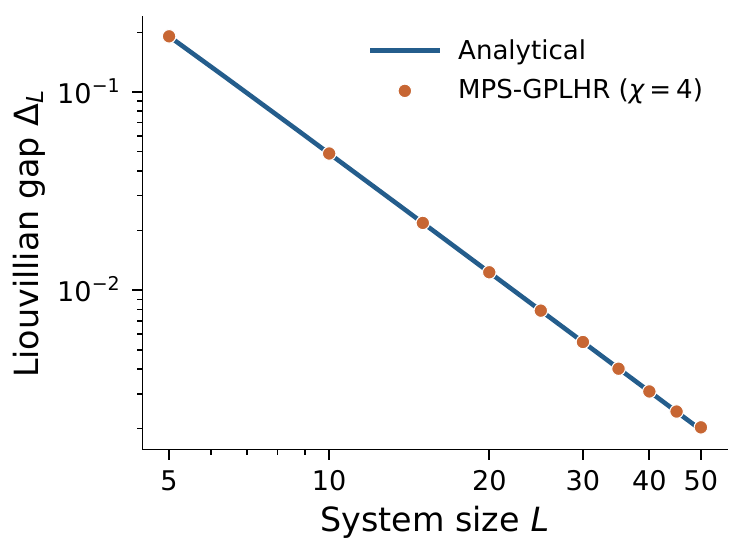}
  \end{minipage}
  \caption{Low-lying spectrum and gap of the periodic quadratic fermionic Liouvillian for $J=0.25$, $\kappa=1$, and $\phi=\pi/L$, computed using
  MPS-GPLHR with $D=4$.
  (a) Five computed eigenvalue estimates at $L=20$.
  (b) Gap estimates for $L=5,10,\ldots,50$ (orange points), compared with
  the exact expression $2\kappa\sin^2[\pi/(2L)]$ (blue curve).}
  \label{fig:quadratic-spectrum-gap}
\end{figure*}
We then re-group each site's ket and bra occupation indices into $p_j=2n_j+m_j\in\{0,1,2,3\}$, giving a chain with local dimension 4. The correspondingly permuted matrix is represented as an MPO of bond dimension 15, including the Jordan-Wigner signs in the periodic terms.
Our targets are the five right eigenoperators that are nearest to the origin,
$\mathcal{L}(R_\alpha)=\lambda_\alpha R_\alpha$, equivalently
$\mathbb{L}|R_\alpha\rangle\!\rangle
=\lambda_\alpha|R_\alpha\rangle\!\rangle$.
The zero eigenvalue corresponds to the unique vacuum steady state,
$\rho_{\mathrm{ss}}=|0\rangle\langle0|$, while the real and imaginary parts of the nonzero eigenvalues determine decay rates and oscillation frequencies,
respectively. In particular, the spectral gap
\begin{equation}
  \Delta_L=-\max_{\lambda_\alpha\ne0}\operatorname{Re}\lambda_\alpha
\end{equation}
sets the slowest exponential decay rate.
The origin of the degeneracy is explicit in this model. With
$c_k=L^{-1/2}\sum_j e^{-ikj}c_j$ and $k=2\pi n/L$, the coherence
$|k\rangle\langle0|$, where $|k\rangle=c_k^\dagger|0\rangle$, has
eigenvalue $-(\kappa+2iJ)[1-\cos(k-\phi)]$; its adjoint has the
complex-conjugate eigenvalue. At $\phi=\pi/L$, the two distinct momenta $k=0$ and $k=2\pi/L$ have the same smallest decay rate. Consequently, the four slowest nonzero eigenoperators form two doubly degenerate eigenvalues,
\begin{equation}
  \lambda_\pm=-\Delta_L\pm i\frac{2J}{\kappa}\Delta_L,
  \qquad
  \Delta_L=2\kappa\sin^2\!\left(\frac{\pi}{2L}\right)
  \sim\frac{\kappa\pi^2}{2L^2}.
  \label{eq:quadratic-exact-gap}
\end{equation}
Thus the characteristic decay time grows as $L^2$, making the low-lying spectrum a useful test as the gap closes with increasing system size.

The initial block includes the vectorized vacuum operator
$|\rho_{\mathrm{vac}}\rangle\rangle$, with
$\rho_{\mathrm{vac}}=|0\rangle\langle0|$,
represented as an MPS of bond dimension one.
We choose the target shift
\begin{equation}
    \sigma=0.1\Delta_L.
\end{equation}
This small positive real shift targets the spectrum near
the origin while avoiding the singular shifted operator
at $\sigma=0$.

Fig.~\ref{fig:quadratic-spectrum-gap} compares the computed spectrum and gap with these expectations. Panel (a) displays the converged $L=20$ spectrum: the four nonzero modes overlap pairwise at the two expected complex eigenvalues, alongside the steady-state root at the origin. Panel (b) shows the gap estimates for $L=5,10,\ldots,50$, demonstrating accurate convergence at $D=4$ in this range. Both doubly degenerate nonzero eigenvalues are recovered, with two linearly independent eigenoperators for each eigenvalue.

\section{Discussion and conclusion}
\label{subsec:conclusion}

In this work, we have developed an MPS formulation of GPLHR for computing multiple right eigenstates of non-Hermitian many-body operators whose eigenvalues lie closest to a prescribed shift $\sigma$ in the complex plane. The method enables efficient spectral targeting without requiring exact shift-and-invert solves, while SRC compression is further employed to reduce the cost of MPO-MPS products.

Benchmarks on transcorrelated and genuinely complex Hubbard chains, the interacting non-Hermitian SSH chain, and a Liouvillian model demonstrate the applicability of MPS-GPLHR to non-Hermitian problems with both real and complex spectra. These results establish MPS-GPLHR as a practical approach to multiple-state calculations in non-Hermitian many-body systems. More effective preconditioners provide a natural direction for further improving its efficiency and extending its range of applicability. In addition, the block-iteration structure of GPLHR is well suited to parallel computation, since operations on multiple trial vectors can be performed concurrently, offering a promising route toward improved scalability for larger many-body problems.

More broadly, MPS-GPLHR belongs to a growing family of algorithms formulated directly in the MPS representation, including MPS-based Lanczos methods, global Krylov evolution, shift-and-invert eigensolvers, and GMRES~\cite{DargelEtAl2012,Paeckel2019,Yu2017,WangEtAl2025,Paeckel2025BandLanczos,Dolgov2013,DektorEtAl2026}.  Developments along these directions could further enhance the efficiency and versatility of MPS-based numerical linear-algebra methods and extend their reach to larger and more challenging many-body problems.
\begin{acknowledgments}
The authors gratefully acknowledge the High Performance Computing Center at The Chinese University of Hong Kong, Shenzhen, for providing computational resources. 
\end{acknowledgments}

\section*{Data and code availability}
The data and code supporting this study are available from the corresponding authors upon reasonable request.

\section*{Author contributions}
X.W. implemented the MPS-GPLHR code,  performed the numerical experiments, and prepared the manuscript. Z.Y. developed the central methodological idea, implemented the MPS-GPLHR code, and revised the manuscript. Y.C. contributed to the discussions. Y.W. contributed to the discussions, co-supervised the project, and participated in writing the manuscript. F.C. supervised the project and revised the manuscript.

\bibliography{references}

\end{document}